# Augmenting the fine beam tube: From hybrid measurements to magnetic field visualization

O. Bodensiek,[1] D. Sonntag,[1] N. Wendorff,[2] G. Albuquerque,[2] and M. Magnor[2]

[1] *Institute for Science Education Research, Physics Group, Technische Universität Braunschweig*
[2] *Computer Graphics Lab, Technische Universität Braunschweig*

Since the emergence of Augmented Reality (AR), it has been a constant subject of educational research, as it can improve conceptual understanding and generally promote learning [1]. In addition, a motivational effect and improved interaction and collaboration through AR were observed [2]. Recently, AR technologies have taken a major leap forward in development, such that especially head-mounted devices or smartglasses came in question for supporting experimentation in STEM education [3,4]. In line with these developments, we here present an AR experiment in electrodynamics for undergraduate laboratory courses in physics using real-time physical data from and virtual tools on mobile devices to both analyze and visualize physical phenomena.

## THEORETICAL BACKGROUND

In order to determine the electron charge-to-mass ratio $-e/m_e$ a fine beam tube is typically used in educational settings. Its main part is an electron gun that generates electrons by thermal glow emission, accelerates them due to a voltage $U_{acc}$ between anode and cathode and bundles them into a focused beam. The electron gun is embedded into an evacuated glass sphere back-filled with hydrogen or helium at low-pressure. Hence, the electron beam becomes visible due to impact ionization. This fine beam tube is mounted on a stand right in the middle of a Helmholtz coil pair, where a coil current $I_{coil}$ generates an almost homogenous magnetic field. Provided $I_{coil}$ is large enough compared to $U_{acc}$, the resulting Lorentz force deflects the electrons onto a circular path within the glass sphere. The Lorentz force acting on an electron with velocity $\vec{v}_e$ in a magnetic field $\vec{B}$ is given by

$$\vec{F} = -e\left(\vec{v}_e \times \vec{B}\right).$$

For an ideal homogenous magnetic field and the electron velocity being perpendicular to the magnetic field, the Lorentz force acts as radial force, that is,

$$m_e \cdot \frac{v_e^2}{r} = -e \cdot v_e \cdot B$$

according to amount and with $r$ representing the radius of the circular electron path. Using the relations between $v_e$ and $U_{acc}$ respectively between $B$ and $I_{coil}$ the essential proportionality in this experiment is given by

$$-\frac{e}{m_e} \propto \frac{U_{acc}}{r^2 \cdot I_{coil}^2}.$$

## EXPERIMENTAL SETUP

In the experiment students use the Microsoft HoloLens [5] as AR smartglasses both to record measurement data and to study the physics of charged particles in magnetic fields in a hybrid, i.e. digitally enhanced, lab environment.

$U_{acc}$ and $I_{coil}$ are measured with multimeters and gathered by a USB-connected single board computer [cf. Fig. 1], which in turn establishes a wireless data link to the HoloLens. On the HoloLens' semi-transparent display the real-time measurement data is presented as numerical values in real time [cf. Fig. 2]. In order to determine the radius of the electron beam, we have added a virtual ruler that can by gesture control both be moved in depth to the plane of the electron beam and adjusted to the diameter of the circular beam path [cf. Fig. 2].

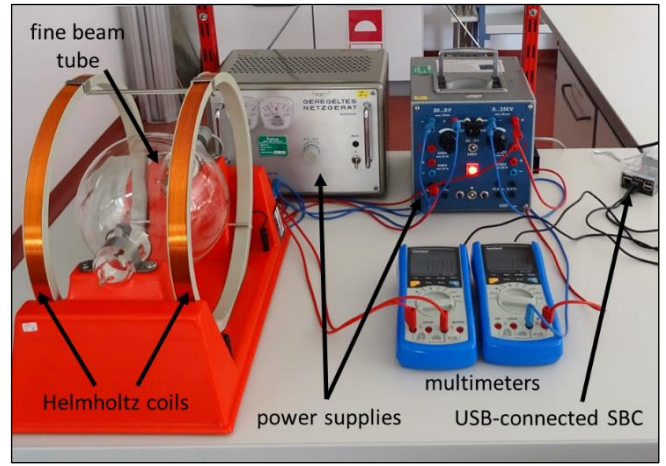

Fig. 1. Experimental setup of the fine beam tube with Helmholtz coils, power supplies and multimeters connected via USB with a single board computer (SBC).

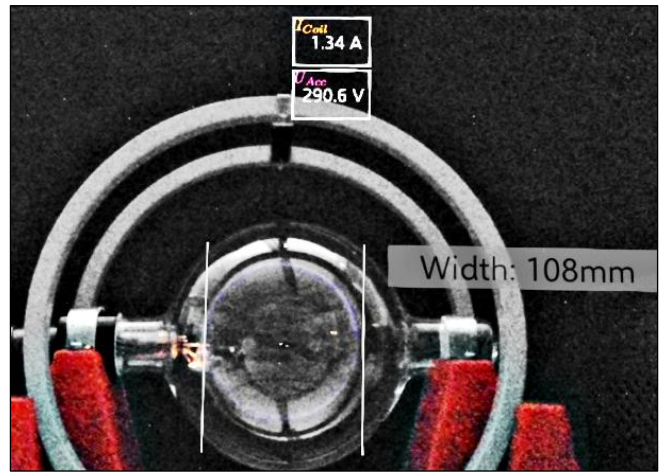

Fig. 2. View through the smartglasses while using the virtual ruler.

All numerical values that are needed to calculate the charge-to-mass ratio are shown on the display of the smartglasses. By "air tapping" a single record button (out of the field of view in Fig. 2), all three values are automatically added to a CSV file that can be analysed after the experiment. Additional measurements are recorded either for different voltages and coil currents measuring the altered diameter again, or by adjusting $U_{acc}$ and $I_{coil}$ such that the diameter keeps constant. Especially in the latter case, a measurement series can be done both rather quickly and by a single student, if wanted or necessary.

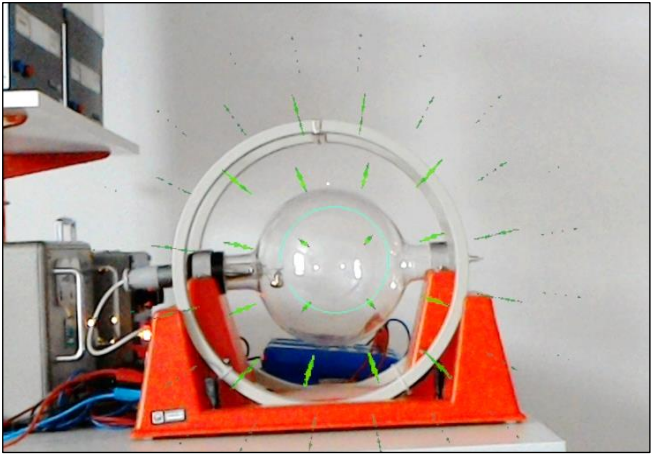

Fig. 3. Visualization of the magnetic field as vector plot. In addition, the theoretically predicted electron beam is augmented.

In addition to the hybrid measurement possibility we implemented a visualization of the magnetic field [cf. Fig. 3] according to the current experimental parameters. The user can also overlay the corresponding theoretically predicted electron beam. All field and beam data is calculated in advance by the finite-element method for a relevant parameter space of $U_{acc}$ and $I_{coil}$ with a sufficiently narrow parameter grid. In between these points on the grid the AR applications interpolates the pre-calculated field and beam data linearly and visualizes it on the smartglasses accordingly. Moreover, the relevant formulas for Lorentz force, field strength as function of $I_{coil}$ and electron velocity as function of $U_{acc}$ can be enabled as overlay to the experiment (not shown here). The parameter(s) currently being changed in the experiment are then highlighted by color in the formulas.

## EXPERIMENTAL RESULTS

We recorded a series of N=18 measurements with constant radius but different acceleration voltages and coil currents in the AR environment. As a mean value we obtained $-e/m_e = -(1.76 \pm 0.01) \cdot 10^{11}$ C·kg$^{-1}$, which is remarkably close to the CODATA value [6] $-1.758820024(11) \cdot 10^{11}$ C·kg$^{-1}$. In several manual measurement series, visually reading off the values, we best reached an accuracy of only about 3%. We relate the improved accuracy in the AR environment to the following two factors: One the one hand, even the stabilized power and voltage supplies we used vary over time. As an effect, one or both values read off may differ a little from the ones immediately after calibration to the constant radius. This error source is reduced in our AR measurement, where both values are recorded simultaneously right after calibration. On the other hand, visually reading off the diameter includes a parallax error, as the measuring device lies approximately 9 cm in front of the electron beam. This error can be reduced by using a mirror on the backside of the glass sphere but it is still not as accurate as the virtual ruler in our AR environment, which can be placed directly in the plane of the electron beam in the glass sphere.

The only drawback we experienced in using the AR environment for this specific experiment is the need to adjust the light conditions so that both the weakly glowing electron beam is clearly visible through the darkening HoloLens and one can still see enough to operate the power supplies.

## CONCLUSION

We implemented an augmented lab experiment for the fine beam tube using AR smartglasses. All measurements in order to determine the specific charge can be recorded digitally in the AR environment. With this AR-based approach we observe several advantages: First, the measured values seem to be more accurate compared to reading them off visually. Second, acquisition of measurement values is easy and quick and can easily be done alone. Finally and probably most important, the additional field visualization coupled to real-time data provides an immediate feedback to the students' experimental actions. In combination with corresponding mathematical formulas of a theoretical description in a single hybrid learning environment we expect this to foster understanding relationships between theory and experiment as found in comparable AR experiments [4] since it provides high temporal and spatial contiguity thereby avoiding a split-attention effect [7].